\documentclass[journal=jacsat,manuscript=article]{achemso}

\usepackage[version=3]{mhchem}

\author{Hooman Barati Sedeh}
\author{Mohammad Hosein Fakheri}
\author{Ali Abdolali}
\email{Abdolali@iust.ac.ir}
\affiliation{
	Applied Electromagnetic Laboratory, School of Electrical Engineering, Iran University of Science and Technology, Tehran, 1684613114, Iran}
	\author{Fei Sun}
\affiliation{
	College of Physics and Optoelectronics, Taiyuan University of Technology, Taiyuan, 030024 China.}
\title[An \textsf{achemso} demo]
  {Thermal-null medium (TNM) : a novel material to achieve feasible thermodynamics devices beyond conventional challenges}
\begin{document}
\begin{abstract}
Recently, heat manipulation has gained the attention of scientific community due to its several applications. In this letter, based on transformation thermodynamic (TT) methodology, a novel material, which is called thermal-null medium (TNM), is proposed that enables us to design various thermal functionalities such as thermal bending devices,  arbitrary shape heat concentrators and omnidirectional thermal cloaks. In contrary to the conventional TT-based conductivities, which are inhomogeneous and anisotropic, TNMs are homogeneous and easy to realize. In addition, the attained TNMs are independent of the device shape. That is if the geometry of the desired device is changed, there is no need to recalculate the necessitating conductivities. This feature of TNM will make it suitable for scenarios where re-configurability is of utmost importance. Several numerical simulations are carried out to demonstrate the TNM capability and its applications in directional bending devices, heat concentrators and thermal cloaks. The proposed TNM could open a new
avenue for potential applications in solar thermal panels and thermal-electric devices.

\end{abstract}
\section{1. Introduction}
In the times of energy shortage, the recycling of heat energy and heat manipulation becomes an important topic. Therefore, how to handle the
heat dissipation, the heat storage and the control of heat flux becomes a significant subject of debates among the scientific community. Various methods have been proposed for this aim which among all of them, transformation thermodynamics (TT) demonstrates a high flexibility to manipulate heat in an unprecedented manner \cite{guenneau2012transformation,schittny2013experiments}. 
The main idea of TT is derived from its electromagnetic (EM) counterpart which was proposed by Pendry \textit{et al.} and named as transformation optics (TO) \cite{pendry2006controlling}. As it was analyzed by Pendry's group, an equivalence between Maxwell's
equations described in an initial coordinate system (i.e. virtual
domain) and their counterparts in another arbitrary transformed
coordinate system (i.e. physical domain), will result in a direct link
between  permittivity and permeability of the occupying
material and the transformed space metric tensor, which contains
the desired EM properties. Soon after the introduction of TO, many novel devices which deemed impossible to be achieved with natural materials such as EM cloaks \cite{liu2009broadband,xu2015conformal,fakheri2017carpet}, multi emission lenses \cite{tichit2014spiral, zhang2016experimental,jiang2012broadband,ashrafian2019space,barati2019exploiting}, EM wave concentrator \cite{rahm2008design,yang2009metamaterial,sadeghi2015transformation,zhao2018feasible,yang2019arbitrarily,zhou2018perfect,abdolali2019geometry} and beam splitters \cite{barati2018experimental,sedeh2019advanced,kwon2008polarization} have been proposed.
The same as TO methodology, TT has gained lots of attention due to the intrinsic degree of freedom that offers.  Although TT could paves the way towards designing various devices such as heat cloak \cite{guenneau2012transformation}, heat flux concentrators  \cite{yu2011design} and heat transferring devices \cite{hu2016directional}, it has some serious drawbacks which restrict the usage of this approach for real-life scenarios. The main problem of TT is that the attained conductivities through this approach are inhomogeneous and anisotropic which result in difficulties for their fabrications \cite{guenneau2012transformation}. Although thermal metamaterials have received lots of attention in recent years, it is still a challenging task to design a thermal cell which possess both inhomogenity and anisotropy \cite{farhat2016transformation}. In addition to the inhomogenity problem, the TT-based conductivities are extremely depended on the device input shape. That is if the geometry of the device is changed, one must not only perform tedious mathematical calculations to attain  desired conductivities but also redesign them, which is time consuming and not applicable for scenarios where re-configurability is important. Therefore, the above-mentioned reasons cause this question to be asked that whether is there any alternative way of designing heat devices that its necessitating conductivities are homogeneous and independent of the input geometry?\par
In this paper, we propose a new method for achieving various thermal devices based on new material called thermal-null medium (TNM) \cite{sun2019thermal,liu2018fast}. In contrary to the conventional TT-based conductivities, the attained TNMs through our approach are homogenous and geometry free. That is if the desired device shape is changed, there is no need to recalculate the necessitating conductivities. This will make the TNM a good alternative for being used in scenarios where re-configurability is of utmost importance. For instance, as the first application of TNM, a directional heat bending device is designed that regardless of its deflection angle, only one material is used and perfect functionality will be achieved. In addition, for the first time we have used the TNM as a mean to obtain arbitrary shape heat concentrators with homogenous materials. It is shown that by utilizing a TNM, one can freely change the geometry of the concentrator while using the same material. Finally, a thermal omnidirectional cloak is also proposed that is capable of guiding the thermal distributions from any incident direction in a manner that they become undetectable from an outside observer.   \par 
The rest of this paper is organized as follows. First, the fundamental theoretical formulation of the method is given, then several
numerical simulations are performed to demonstrate the capability of propounded approach for achieving different practical functionalities in section 3. Finally, the paper will end with a summary of the
obtained results in the conclusion section.

\section{2. Theoretical formulations}
The derivation of TNM is based on the null-space mapping, which is a kind of transformation that will result in optical/thermal materials with large anisotropic parameters. Here, we will briefly discussed the derivation approach, while for more comprehensive study one can refer to \cite{sun2019thermal,liu2018fast}. Assume a slab with width of $W_v$ is  mapped to another slab with width of $W_p$ as shown in Fig.1.
\begin{figure}[h]
	\centering
	\includegraphics[width=0.6\linewidth]{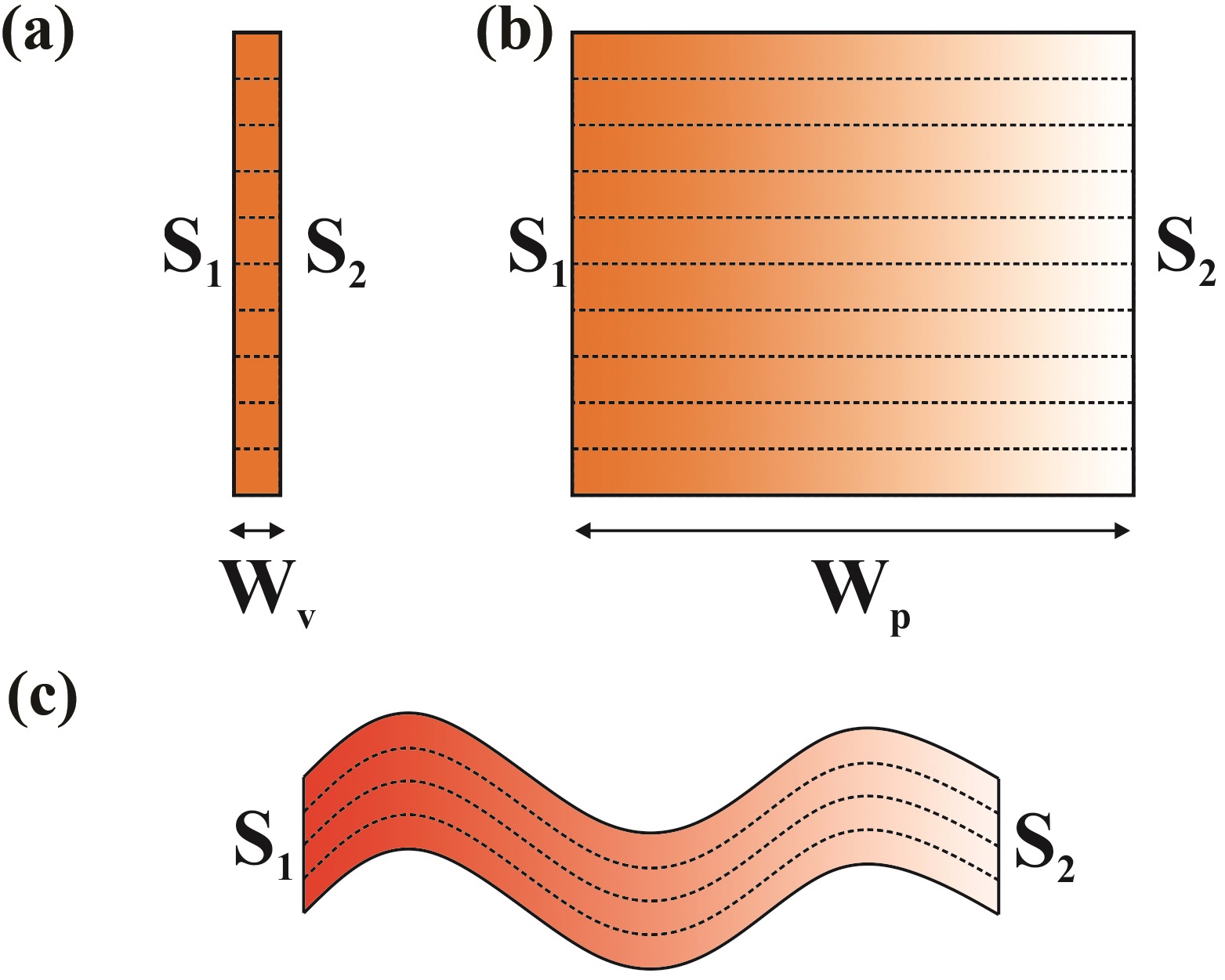}
	\caption{Schematic diagram of null-space transformation to obtain TNM. a)
		A thin slab in the virtual space is mapped to b) a slab with finite thickness. c) Schematic of a typical shape designed by bending a slab in (b).}
	\label{fgr:fig1}
\end{figure} 
The transformation function of such a mapping could be written as
\begin{align}
x^\prime =
\   \left\{
\begin{array}{lll}
x & x^\prime \in (-\infty, 0]\\
\frac{W_p}{W_v} x  &  x^\prime \in [0, W_p] \\
x-W_v+W_p &  x^\prime \in [W_p,\infty) 
\end{array} 
\right. \
\end{align}
When the thickness of the slab in the virtual space approaches zero (i.e.,  $W_v \rightarrow 0$),  its interfaces become very close to each other (see $S_1$ and $S_2$ in Fig.1 (a)).  After a null-space
transformation (i.e., Eq. (1) when $W_v \rightarrow 0$), the thickness of this slab is greatly expanded. In other words, while surface $S_1$ is fixed, surface $S_2$ is stretched by $W_p/W_v$ times in the
physical space as shown in Fig.1 (b). For a steady state case, the thermal conduction
equation without a source can be written as $\nabla\cdot(\kappa \nabla T)=0$, where $\kappa$ is  thermal conductivity and $T$ is the temperature. On the basis of the
invariance of heat conduction equation under coordinate
transformations \cite{guenneau2012transformation}, the thermal conduction equation in the
transformed space can also be written as $\nabla^\prime \cdot(\kappa^\prime \nabla^\prime T^\prime)=0$, through
which we can obtain
\begin{equation}
\kappa^\prime=\frac{\Lambda \times \kappa \times \Lambda^T}{det(\Lambda)}
\end{equation}
where $\Lambda=\partial(x^\prime,y^\prime,z^\prime)/\partial(x,y,z)$ is the Jacobian matrix which relates the metrics of virtual space ,$(x,y,z)$, to the ones of physical space, $(x^\prime,y^\prime,z^\prime)$. By substituting Eq. (1) into Eq. (2) the corresponding heat conductivity will be achieved as
\begin{equation}
\frac{\kappa^\prime}{\kappa}=
\begin{bmatrix}
\frac{W_p}{W_v} &
0&
0 \\
0 &

\frac{W_v}{W_p} &
0\\
0 &
0 &
\frac{W_v}{W_p}
\end{bmatrix} 
\end{equation} 
However, since $W_v \rightarrow 0$ (null-space transformation), the obtained conductivity will be changed to
\begin{equation}
\kappa^\prime=
\begin{bmatrix}
\infty &
0&
0 \\
0 &

0 &
0\\
0 &
0 &
0
\end{bmatrix} 
\end{equation} 
The obtained material of Eq.(4) is named as thermal null medium (TNM), which was designed by null-space
transformation. Since thermal resistance is inversely
proportional to the thermal conductivity, it could be concluded that $R_\parallel \rightarrow 0$ and $R_\perp \rightarrow \infty$, which leads to a super-transferring performance along the “stretching” direction without dissipation. In other words, the function of TNM is to guide heat very fast along its main stretching direction (i.e., $\hat{x}$) and extremely slow in other directions (i.e., $\hat{y}$ and $\hat{z}$ ).  In addition, as it was comprehensively discussed recently in \cite{sun2019thermal,liu2018fast}, TNMs are not restricted to the cartesian coordinate system. In fact, TNMs with any other shapes could be obtained by bending the slab of Fig.1(b) and still guide the thermal field directionally without any dissipation as shown in Fig.1 (c).  \par
\section{3. Numerical simulations}
To verify the correctness of the proposed TNM, heat directional bending devices, are analyzed by performing simulations that were carried out using COMSOL Multiphysics finite element solver. Fig. 2 shows the primary goal of our design. As it can be seen, there is a need to change the direction of a heat flow to the angle of $\alpha$. \par
\begin{figure}[!h]
	\centering
	\includegraphics[width=0.5\linewidth]{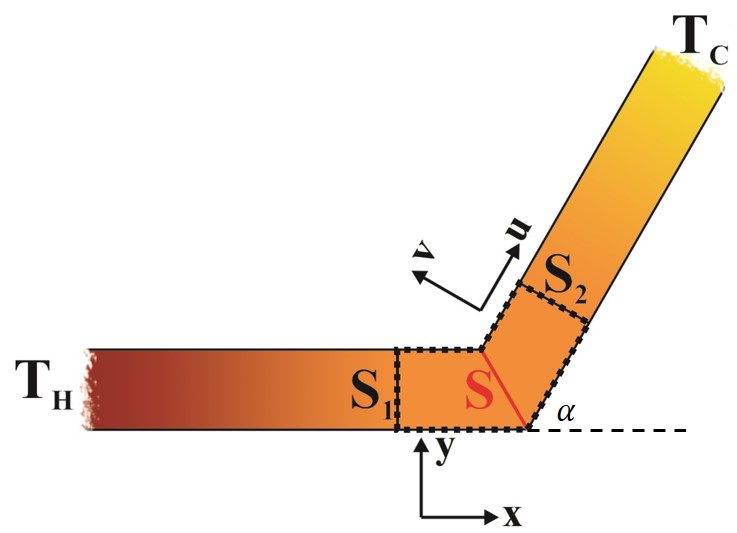}
	\caption{The schematic transformation for describing heat directional bending device }
	\label{fgr:fig4}
\end{figure} 
As it is shown in Fig.2,  the hot interface is separated from the cold one with the deflection angle of $\alpha$. The output surface (that has higher temperature $T_H$) and input surface (that has lower temperature $T_C$) are labeled by $S_1$ and $S_2$,
respectively. It is clear from Fig.2 that a common plane of $S$ (shown with red color) can be obtained, if one extends the edges of surfaces $S_1$ and $S_2$ (indicated by black dashed lines). According to the above-mentioned discussion, filling the region between surfaces $S_1$ and $S$ with a TNM with its main axis along the $\hat{ x}$ direction (i.e., $\kappa_x \rightarrow \infty$ and $\kappa_y \rightarrow 0$) and the area between surfaces $S$ and $S_2$ with a TNM with its main axis along
along the \textit{u} direction (which is rotated to the angle of $\alpha$ with respect
to the $\hat{x}$ direction), will result in point-to-point mapping and in turn will give rise to bending the heat distributions directionally from $S_1$ to $S_2$. In other words, surfaces of $S_1$, $S_2$ and $S$ are all equivalent surfaces, in a manner that heat distributions
on $S_1$ will first be projected onto $S$ along the $\hat{x}$ direction and
then projected onto $S_2$ along the $\alpha$ direction. It is notable to mention that the entire design process is general, without any mathematical calculations and is valid for any arbitrary deflection angle of $\alpha$. To demonstrate such a capability, several thermal bending devices with different bending angles and structures are simulated and their results are depicted in Fig.3. As can be seen from this figure, the obtained TNM is capable of connecting two thermal surfaces with any desired temperature without any deviation in their thermal contours. This will make it a suitable material for being used in scenarios where reconfigurability is of utmost importance or there is a prerequisite to manipulate heat distributions directionally in an arbitrary path. Furthermore, according to Fig.3 (d), the fictitious surface $S$ could also be located in cylindrical coordinate system (i.e., azimuthal direction ($\hat{\phi}$) in Fig.3 (d)) rather than cartesian coordinate and remains its functionality. \par 
\begin{figure}[t]
	\centering
	\includegraphics[width=0.8\linewidth]{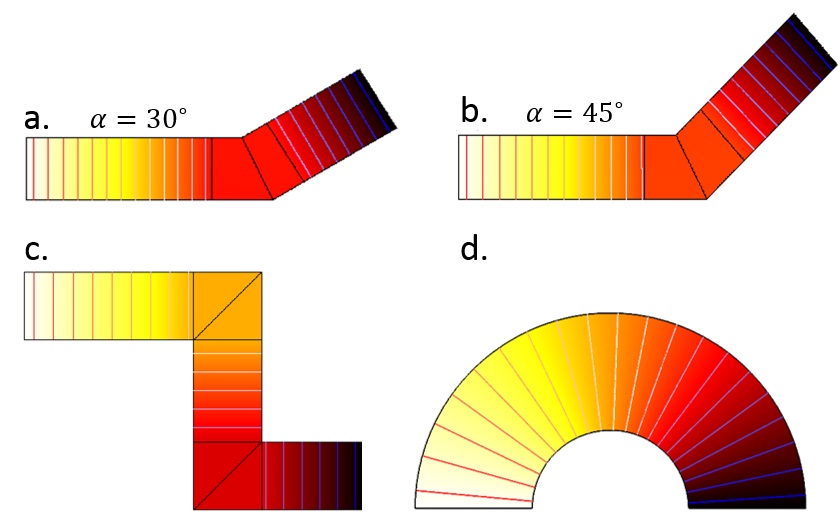}
	\caption{Thermal distributions of a heat bending device with different bending angles of (a) $\alpha=30^\circ$,(b) $\alpha=45^\circ$ and with TNM in  (c) cartesian, (d) cylindrical coordinate system}
	\label{fgr:fig7}
\end{figure} 
In addition to the thermal directional bending devices, over the last few decades, much more attention has been paid to renewable energy since the fossil fuels are being depleted. Among all the proposed methods, solar energy has been widely exploited as one of the most important renewable energy for converting the sun energy to electricity. In general, the conversion of solar energy into electricity is performed via two approaches of direct solar-electrical
energy conversion and indirect conversion. In the former, the solar cell devices are used while in the latter 
thermal energy is used as a mediator by means of a device named as solar thermal collector (STC). Heat flux concentrator is an example of the second group (i.e., STC) that received a lot of attention in the past decade. Recently, TT method has also been extended to manipulate heat current and localize thermal energies with heat flux concentrators. However, the conventional challenges of TT-based conductivities (i.e., inhomogenity and shape dependent) are still the main drawback of TT-based concentrators and restrict their applicability for being used in practical situations. However, by extending the idea of null-space transformation, one can exploit TNMs as a mean to collect thermal energies in any arbitrary region of interest.To this aim, the schematic diagram of Fig.4 is used as the space transformation, which three cylinders with arbitrary cross sections of $R_1(\phi)= \tau_1 R(\phi)$, $R_2(\phi)=\tau_2 R(\phi)$ and $R_3(\phi)=\tau_3 R(\phi) $ divide the space (i.e., virtual space) into three different regions. It should be noted that  $\tau_1$, $\tau_2$ and $\tau_3$ are constant coefficients which satisfy the condition of $\tau_1 < \tau_2 < \tau_3$  and $R(\phi)$ is an arbitrary continuous function with period of $2\pi$  that is specified with Fourier series as
\begin{equation}
R(\phi) = a_{0}+ \sum_{n=1}^{\infty} \{ a_{n} \cdot cos (n\phi) +b_{n} \cdot cos (n\phi) \}
\end{equation}

where $a_n$ and $b_n$ are constant coefficients that specify the contour shape. \par 
\begin{figure}[!h]
	\centering
	\includegraphics[width=\linewidth]{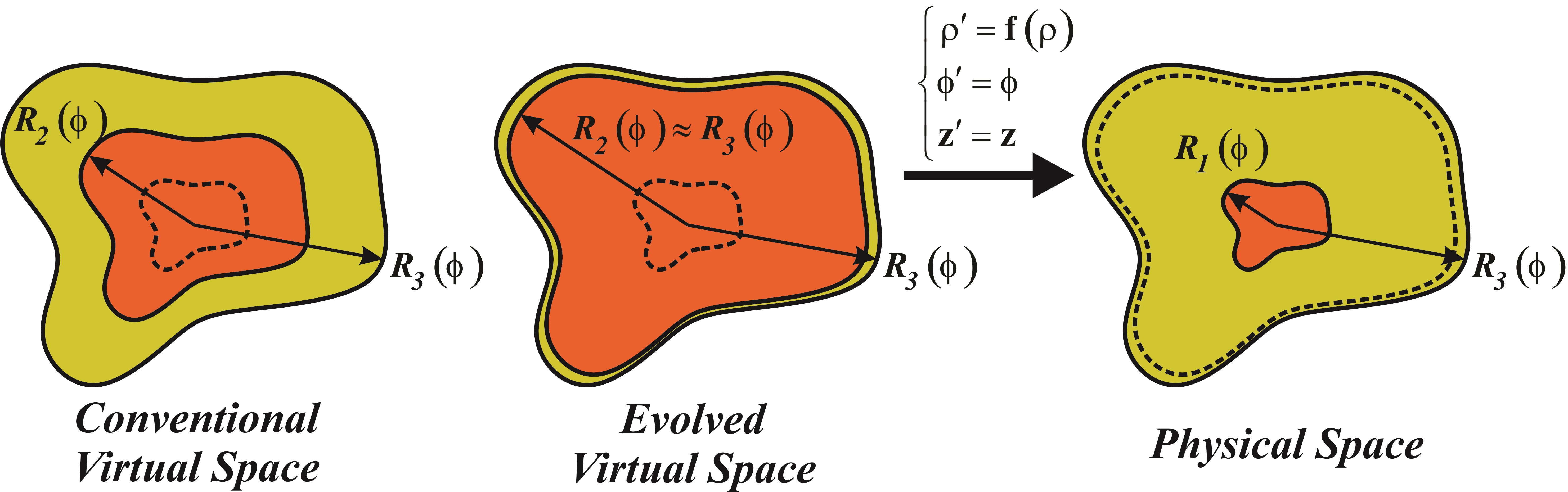}
	\caption{ The schematic of coordinate transformation for achieving arbitrary shape concentrators.}
	\label{fig:fig8}
\end{figure}
To concentrate the thermal energy in a predefined region of $R_1(\phi)$, one must collect the energy that was originally located in $\rho<R_2(\phi)$ into the region of $\rho^{\prime} < R_1 (\phi^\prime)$ as shown in Fig.4. To this aim, as explained in \cite{guenneau2012transformation,farhat2016transformation}, the region of $\rho\in[0, R_2 (\phi)]$ must be compressed into the region $\rho^\prime\in[0, R_1 (\phi^\prime)]$, while at the same time $\rho\in[R_2 (\phi), R_3 (\phi)]$ is stretched into the region of $\rho^\prime\in[R_1 (\phi^\prime), R_3 (\phi^\prime)]$.
Since these two steps occur simultaneously, all the energy, which previously located in $\rho < R_2 (\phi)$ is now localized in the region of $\rho^{\prime} < R_1 (\phi^\prime)$  and as a result, the thermal intensity will be increased in the mentioned domain. The function which is competent to perform such a transformation could be expressed as
\begin{equation}
\   \left\{
\begin{array}{ll}
f_c(\rho,\phi)= \frac{\tau_1}{\tau_2}\rho& \rho'\in [0, R_1(\phi))\vspace{0.5 cm} \\
f_s(\rho,\phi)= \Omega \times \rho + \Psi \times R(\phi)  & \rho'\in[R_1 (\phi), R_3 (\phi)]
\end{array} 
\right. \
\end{equation}
where subscripts of $c$ and $s$ represent the compressed and stretched region, respectively and $\Omega=(\tau_3-\tau_1)/(\tau_3-\tau_2)$, $\Psi= [(\tau_1-\tau_2)/(\tau_3-\tau_2)]\tau_3$. By substituting Eq. (6) into Eq. (2) the necessitating materials for each region will be achieved as
	\begin{equation}
	\frac{\kappa'_{c}}{\kappa_0}=
	\begin{bmatrix}
	1 &
	0&
	0 \\
	0 &
	
	1 &
	0\\
	0 &
	0 &
	(\tau_2/\tau_1)^2
	\end{bmatrix} ,
	\frac{\kappa'_{s}}{\kappa_0}=
	\begin{bmatrix}
	m_{11} &
	m_{12}&
	0 \\
	m_{21} &
	
	m_{22}&
	0\\
	0 &
	0 &
	m_{33}
	\end{bmatrix} 
	\end{equation}
where the coefficients of $m_{ij}$ are
\begin{align}
&m_{11} = \frac{(\tau_3 -\tau_2)\rho^\prime - \tau_3(\tau_1-\tau_2)R(\phi)}{(\tau_3-\tau_2)\rho^\prime}+ 
\frac{\tau^2_3(\tau_1-\tau_2)^2(dR(\phi)/d\phi)^2 }{[(\tau_3-\tau_2)^2 r^\prime - \tau_3 (\tau_1-\tau_2)(\tau_3-\tau_2)R(\phi)]\rho^\prime} \\ \nonumber
&m_{12} = m_{21}= \frac{(\tau_3(\tau_1-\tau_2)(dR/d\phi))}{(\tau_3-\tau_2)\rho^\prime- \tau_3 (\tau_1 -\tau_2) R(\phi)},m_{22} = \frac{(\tau_3-\tau_2)r^\prime}{(\tau_3-\tau_2)\rho^\prime-\tau_3(\tau_1-\tau_2)R(\phi) } \\ \nonumber
&m_{33}=\frac{(\tau_3-\tau_2)^2 \rho^\prime -\tau_3(\tau_1-\tau_2)(\tau_3-\tau_2)R(\phi)}{(\tau_3-\tau_1)^2 \rho^\prime}
\end{align}

As can be seen from Eq.(7) and Eq.(8), the obtained materials of the stretched region are inhomogeneous and anisotropic with off-diagonal components of $m_{12} / m_{21}$ which cause serious difficulties in their realization process. In fact, the reason why there is no experimental verification of arbitrary shape thermal concentrators are yet proposed is due to the existence of these inhomogeneous and off-diagonal components. \par 
In addition to the inhomogenity that the existence of $R(\phi)$ in the components of $m_{ij}$ dictates, it also demonstrate the dependency of obtained materials to the structure geometry. In other words, an alternation in the coefficients of Eq.(5),which results in new geometry, leads to new materials which must be recalculated. However, since $ R_2(\phi)=\tau_2 \times R(\phi)$  is a fictitious region, $\tau_2$  can achieve any arbitrary value. This will give us a degree of freedom to arbitrarily select the value of $\tau_2$ in a manner that it will eradicate the effect of the off-diagonal components of $m_{12}$ (also $m_{21}$). Therefore, without the loss of generality one can assume that $\tau_2 \rightarrow \tau_3$(null-space transformation). By setting such a value for $\tau_2$ the coefficients of Equation (8) will be attained as
\begin{equation}
\   \left\{
\begin{array}{lll}
m_{11}=\frac{1}{\Delta}
\vspace{0.5 cm} \\ 
m_{12}=m_{21}=-\kappa_0\frac{dR(\phi)/d\phi}{R(\phi)}
\vspace{0.5 cm} \\ 
m_{22}=m_{33}=\Delta
\end{array} 
\right. \	
\end{equation}
whence $\Delta \rightarrow 0$. As can be seen from Equation (9), the obtained conductivities are still suffer from anisotropy and inhomogeneity problem due to the existence of $m_{12}$ (also $m_{21}$ ) and their dependence to the contour shape of $R(\phi)$. However, it is known that in the absence of a heat source the heat diffusion equation of steady state is governed by the Laplace equation as 
\begin{align}
&\nabla \cdot(\bar{\bar{\kappa}}\nabla T)=\sum_{j,k=1}^{n}\frac{\partial  ^{j,k}}{\partial x^j}\kappa\frac{\partial T}{\partial x^k}=0
\end{align}
Therefore, after some simple calculation the Laplace equation in cylindrical coordinate will be achieved as
\begin{align}
&\frac{1}{\rho} m_{11} \partial T/\partial \rho + m_{11}  \partial^2 T/\partial \rho^2 + \frac{2}{\rho} m_{12}  \partial^2 T/\partial \rho \partial \phi +
\frac{1}{\rho}  \frac{\partial m_{12}}{\partial \phi}   \partial T/\partial \rho +\frac{1}{\rho^2}m_{22} \partial^2 T/\partial \phi^2 \\ \nonumber
& +m_{33}  \partial^2 T/\partial z^2 =0           
\end{align}
By substituting Equation (9) into Equation (11), one can easily obtain
\begin{align}
&\frac{1}{\rho} \partial T/\partial \rho +  \partial^2 T/\partial \rho^2 + \frac{2}{\rho} m_{12} \times \Delta \times  \partial^2 T/\partial \rho \partial \phi + 
\frac{1}{\rho} \frac{\partial m_{12}}{\partial \phi} \times \Delta \times \partial T/\partial \rho \\ \nonumber & +\frac{1}{\rho^2}\Delta^2 \partial^2 T/\partial \phi^2 
 +\Delta^2  \partial^2 T/\partial z^2 =0           
\end{align}	
Since $\Delta \rightarrow 0$ and $m_{12}$ has a finite value according to Eq. (9), the exact value of  $m_{12}$ is not important.  This is because only the products of these values play crucial role as shown in the Laplace equation (i.e. Eq. (12)) not each of them individually. Therefore, one can assume any desirable finite value for the off-diagonal components. Here we assumed $m_{12}=0$, this assumption would eradicate the off-diagonal components of Eq. (9).  Therefore, the final conductivities, which describe the performance of an arbitrary shape thermal concentrator, will be achieved as
	\begin{equation}
	\frac{\kappa'_{c}}{\kappa_0}= 
	\begin{bmatrix}
	1 &
	0&
	0 \\
	0 &
	
	1&
	0\\
	0 &
	0 &
	(\tau_2/ \tau_1)^2
	\end{bmatrix} ,
	\frac{\kappa'_{s}}{\kappa_0}=
	\begin{bmatrix}
	\infty &
	0&
	0 \\
	0 &
	
	0&
	0\\
	0 &
	0 &
	0
	\end{bmatrix}
	\end{equation}
To validate the concept, several arbitrary shape concentrators were simulated. The solving area is consisted of a square which at the position of $x=-0.3 m$ , there is a planar metallic plate with the temperature  of $T=315 $ K, and the metallic plate at $x=+0.3 m$  has the temperature $T=275$ K. For all the performing simulations, it is assumed that the values of $\tau_1=0.5$, $\tau_2=0.99$ and $\tau_3=1$ are constant while $R(\phi)$ is changed for each new case. Therefore, the necessitating conductivies will be attained according to Eq. (13) as
	\begin{equation}
	\frac{\kappa'_{c}}{\kappa_0}= 
	\begin{bmatrix}
	1 &
	0&
	0 \\
	0 &
	
	1&
	0\\
	0 &
	0 &
	3.92
	\end{bmatrix} ,
	\frac{\kappa'_{s}}{\kappa_0}=
	\begin{bmatrix}
	\infty &
	0&
	0 \\
	0 &
	
	0&
	0\\
	0 &
	0 &
	0
	\end{bmatrix}
	\end{equation}
The first example is dedicated to circular and elliptical cross-section concentrators and their results are illustrated in Figure 5.
\begin{figure}[!h]
	\centering
	\includegraphics[width=0.8\linewidth]{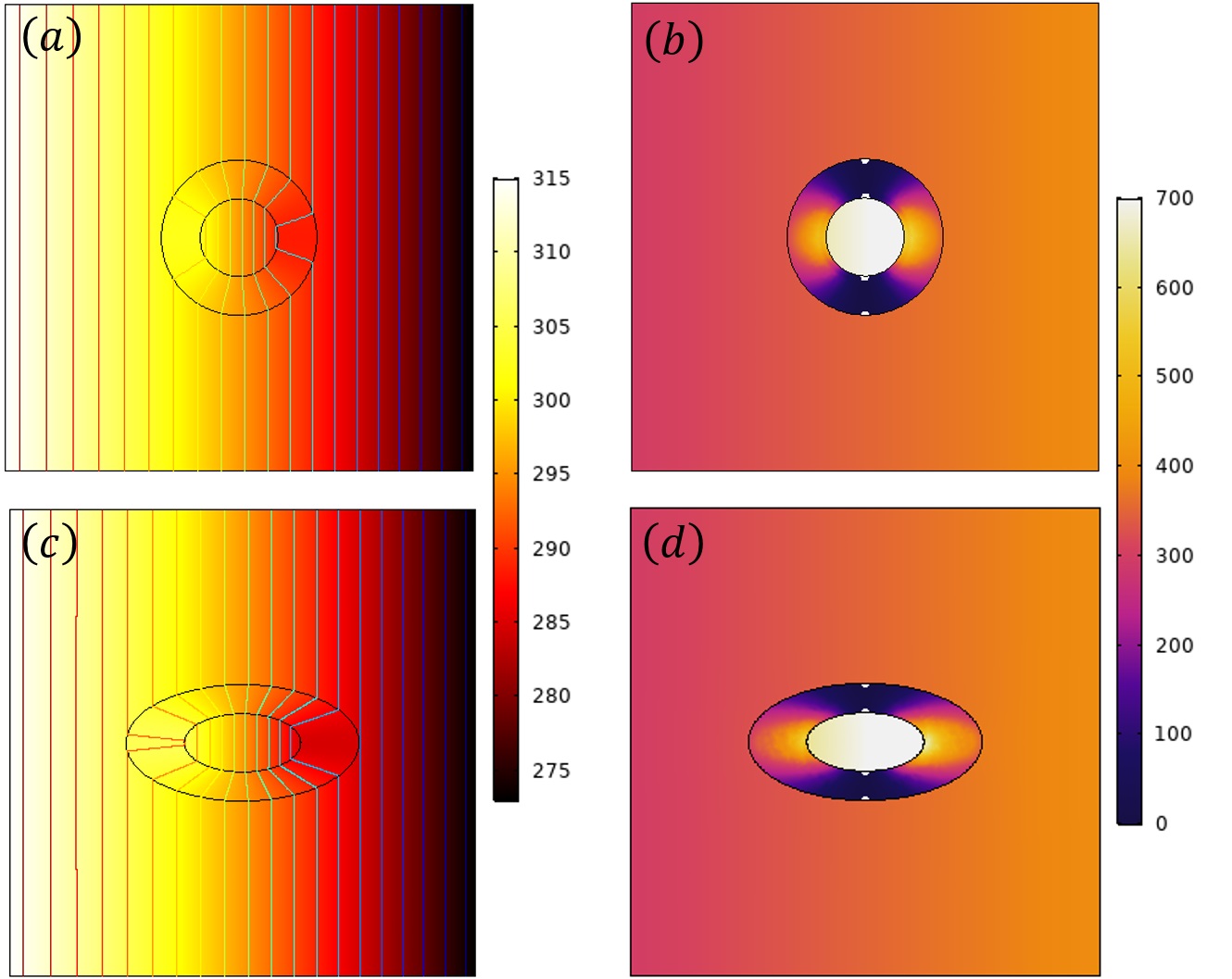}
	\caption{ The temperature distribution results of the heat flux concentrator with different cross sections of (a),(c) circular and elliptical. (b),(d) their corresponding thermal flow.}
	\label{fig:fig9}
\end{figure}
As can be seen from Fig. 5(a) and Fig.5(c), the thermal concentrator does not change the thermal distribution which is well abide with the theoretical investigations. In addition, since in the background medium( in a specific location) the heat flux is equal to $350 w/m^2$, it is expected that in this particular location, the heat flux be enhanced by the ratio of $\tau_2/\tau_1=1.98$. This has been verified by the numerical simulations, as shown in Fig.5(b) and Fig.5(d). It is evident that in the compressed region the heat flux is increased from $350 w/m^2$ to $693 w/m^2$ which is in good agreement with the theoretical predictions.\par 
To have full control of heat flux it might be necessary in some cases that the thermal energy is localized into a certain domain having an arbitrary cross-section. To date, no systematic work has been proposed for achieving arbitrary shape thermal concentrator. As it was mentioned, the presented method in this article will result in anisotropic conductivities that are independent to the geometry of the concentrator. That is the attained conductivites of Eq.(14) are sufficient for any desired geometry. This will in turn make the proposed approach a good candidate for scenarios where reconfigurability is of utmost importance. To show this, assume that the contour coefficients of $R(\phi)$ in Eq. (5) are given in a way that an arbitrary shape concentrator is generated. By utilizing the obtained conductivities (i.e., Eq. (14)) for each compressed and stretched regions, the results of the thermal flux and temperature distributions will be achieved as shown in Figure 6.
\begin{figure}[!h]
	\centering
	\includegraphics[width=0.8\linewidth]{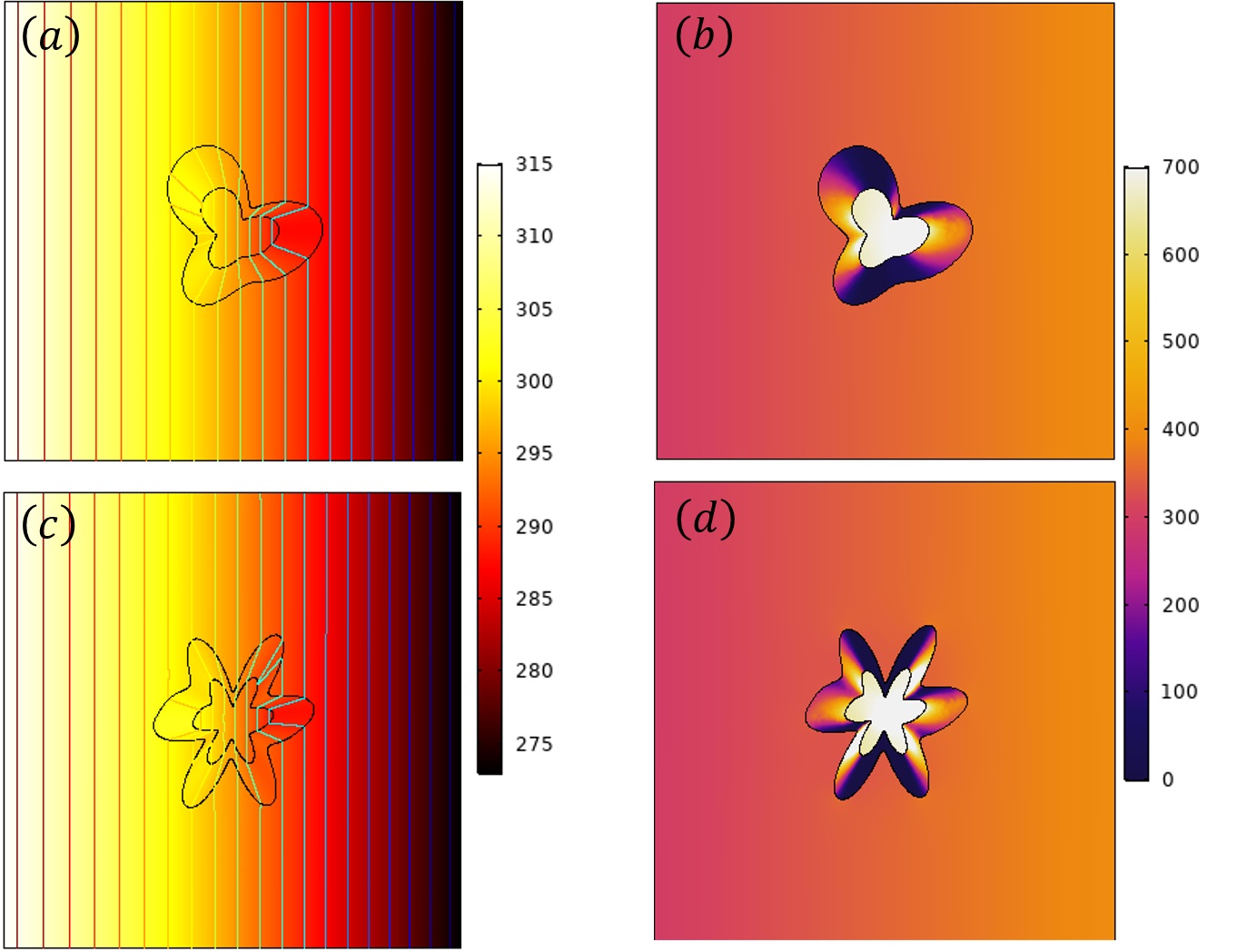}
	\caption{ (a),(c) The temperature distribution results of heat flux concentrator with arbitrary cross-section. (b),(d) their corresponding thermal flow.}
	\label{fig:fig10}
\end{figure}
The same as the previous case, in the new scenario the thermal flow is increased from $350 W/m^2$ to $693 W/m^2$ in the compressed region without any distortion in its temperature distributions. However, it is noteworthy to mention that changing the ratio of $\tau_2/\tau_1$ will result in different values for the thermal energy inside the compressed domain. \par
As the final example, a square shape heat flux cloak is proposed via utilization of TNM. To date, no systematic work has been proposed that yield to omnidirectional heat cloak. However, in this paper for the first time we have proposed a square shape cloak that is capable of guiding the thermal distributions in a manner that the object becomes undetectable from an outside observer. Since heat will be diffused
from a higher temperature region to a lower one, if an obstacle is located in the path of the thermal flux, a distortion in the isothermal lines will occur and in turn give rise to degradation of efficiency. However, when a square shape heat cloak is exploited, it will allow the heat flux to pass smoothly around the cloaked region without creating any distortion as shown schematically in Fig.7 (a). 
\begin{figure}[!h]
	\centering
	\includegraphics[width=0.8\linewidth]{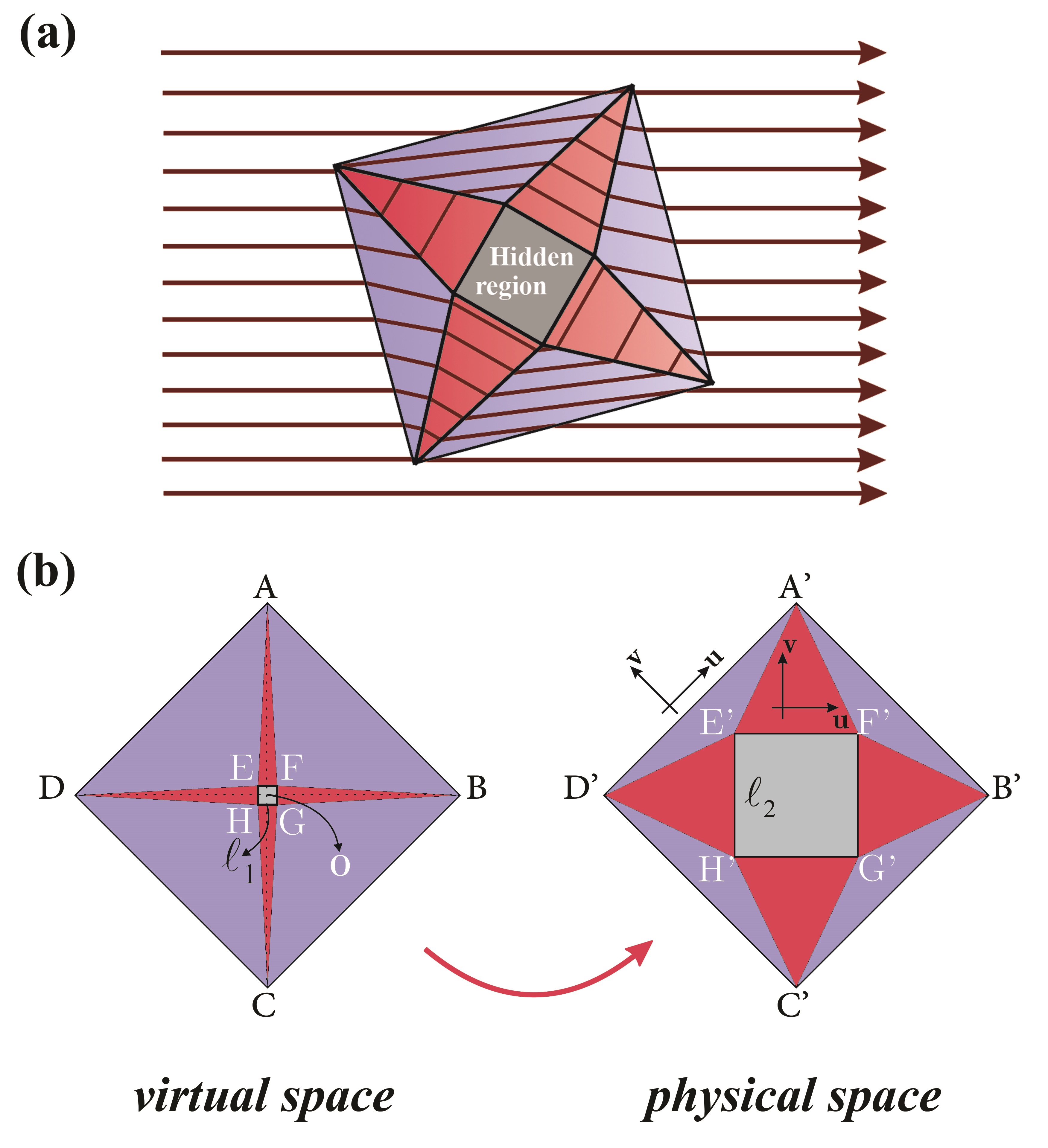}
	\caption{ (a) The schematic of the heat flux path when the TM-based cloak is utilized.(b) The demanding mapping function used for designing a square shape cloak}
	\label{fig:fig11}
\end{figure}
To design such a cloak based on coordinate transformation, the transformation function of Fig.7 (b) is used. In contrary to the previous reported cloaks which mapped a point in the virtual space into a circle in the physical space, here the cloak region has divided into different regions and for each domain a linear transformation function is exploited. The space between the square with side length of $L$ (i.e., \textit{$ A B C D$}) in the virtual space is transformed to the same square with the same length in physical space(i.e., \textit{$A^\prime B^\prime C^\prime D^\prime$}), while at the same time inner square with the side of \textit{$l_1$} (i.e., \textit{$E F G H$})in the virtual space is mapped to a larger square with the side of \textit{$l_2$} (i.e., \textit{$E^\prime F^\prime G^\prime H^\prime$}) in the physical space. \par 
Without the loss of generality, one can assume that the inner square side length is approaching to zero (i.e., $l_1 \rightarrow 0$ ). Therefore, under this assumption, the triangles $\triangle OAB$, $\triangle OBC$, $\triangle OCD$ and $\triangle ODA$ in the virtual
space (Fig.7 (b)) are transformed to triangles $\triangle F^\prime A^\prime B^\prime $, $\triangle G^\prime B^\prime C^\prime$, $\triangle H^\prime C^\prime D^\prime$ and
$\triangle E^\prime D^\prime A^\prime$ in the physical space respectively. Meanwhile, the lines $AO$,
$BO$, $CO$ and $DO$ in the virtual space must also be transformed to
triangles $\triangle A^\prime E^\prime F^\prime$, $\triangle B^\prime F^\prime G^\prime$, $\triangle C^\prime G^\prime H^\prime$ and $\triangle D^\prime H^\prime E^\prime$ in the physical space, respectively. Consequently, the object which was supposed to be cloaked will not affect the heat distributions and will be invisible from the outside detector. By introducing a local coordinate system for each of the pink triangular region ($\triangle D^\prime H^\prime E^\prime$ and $\triangle B^\prime F^\prime G^\prime$ are indicated as I and III, respectively while  $\triangle A^\prime E^\prime F^\prime$ and $\triangle C^\prime G^\prime H^\prime$  are shown by II and IV ), the demanding conductivity for each of these regions is given by 
\begin{equation}
  \frac{\kappa^\prime_{I,III}}{\kappa_0}=
\begin{bmatrix}
0&
0&
0 \\
0 &

\infty &
0\\
0 &
0 &
0 
\end{bmatrix},
 \frac{\kappa^\prime_{II, IV}}{\kappa_0}=
\begin{bmatrix}
\infty &
0&
0 \\
0 &

0  &
0\\
0 &
0 &
0  
\end{bmatrix}
\end{equation}
In addition, the conductivities of the remaining regions (shown by $R$ subscript) in their local coordinate system (i.e.,$ (u,v,z)$) will also be attained as
\begin{equation}
  \frac{\kappa^\prime_R}{\kappa_0}=
\begin{bmatrix}
\frac{1}{\Gamma} &
0&
0 \\
0 &
\Gamma &
0\\
0 &
0 &
\frac{1}{\Gamma}
\end{bmatrix},
\end{equation}
 where $\Gamma=1-(\sqrt{2}l_2/L)$. Compared with the previous cloak prototypes with inhomogeneous and anisotropic extreme material parameters, the square cloak is simplified to one with only two homogeneous materials. The latter is a simple diagonal anisotropic mass density tensor which could be easily implemented via thermal metamaterials \cite{farhat2016transformation,yang2014experimental} and the former is the TNM which is introduced in this paper. To demonstrate the effectiveness of the proposed cloak, we have simulated it with the conductivities given in Eq. (15) and Eq. (16) for each of the corresponding regions under two different angles of $\theta_{inc}=0^\circ$ and $\theta_{inc}=15^\circ$ and their results are shown in Fig.8.
\begin{figure}[!h]
	\centering
	\includegraphics[width=\linewidth]{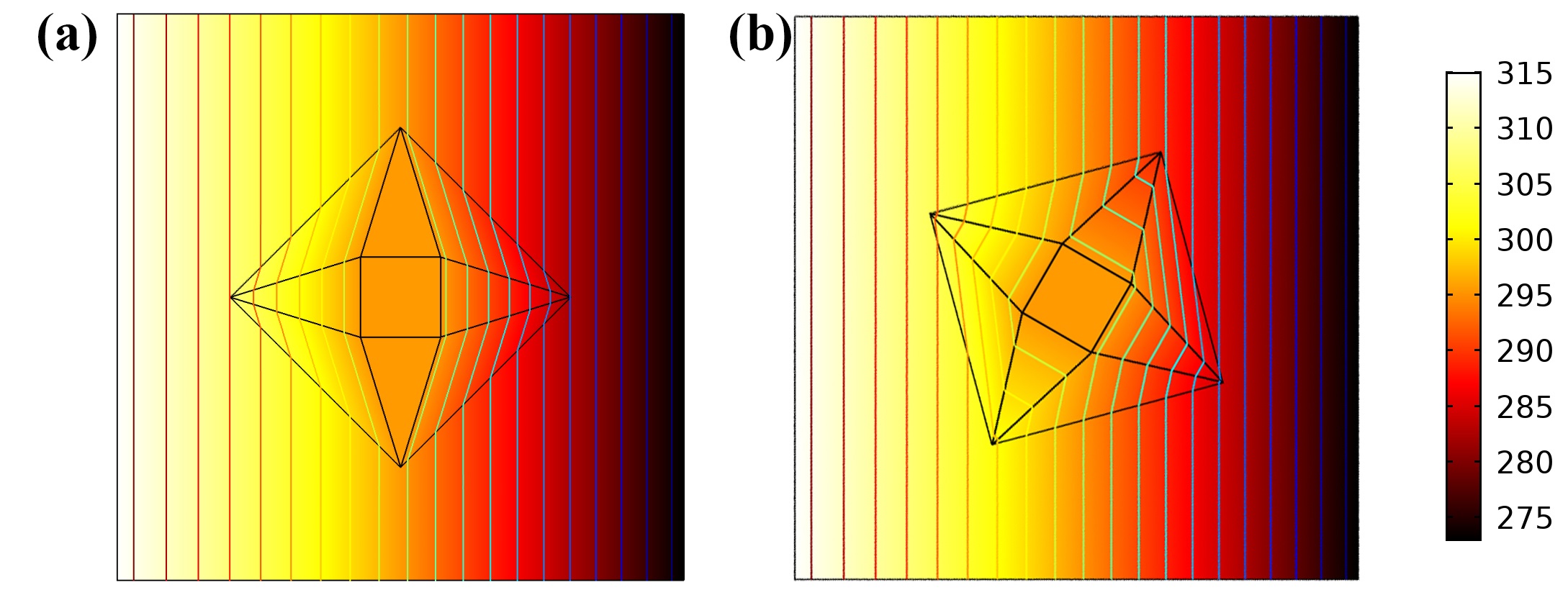}
	\caption{ The heat flux distributions when the designed square shape cloak is used under different incident angles of a) $\theta=0^\circ$ and b) $\theta=15^\circ$}
	\label{fig:fig12}
\end{figure}
As can be seen from Fig.8 (a), the isothermal contours will smoothly pass the object without any distortion. However, in contrary with previous cloaks, when the incident angle is changed, the functionality of the designed cloak will remain unchanged as shown in Fig.8 (b). In other words, the functionality of the proposed cloak is not restricted to any specific incident angle and we used $\theta=15^\circ$ as the demonstration of the concept. This feature of the propounded omnidirectional cloak will consequently give rise to making the object be cloaked from an outside detector no matter where the location of the detector is, which is of utmost importance in many practical applications.
\section{4. Conclusion}
In conclusion, in this paper we have presented a new material based on null-space transformation which is capable of obviating the conventional challenges of TT-based materials. The attained material which is called TNM, is a homogeneous and anisotropic conductivity that is shape independent. In other words, when the geometry of the structure (e.g., the deflection angle of the directional heat devices) is changed, the same material could be used again without the demand of recalculating or re-fabricating it. This will in turn make the proposed material a good candidate to be used for more practical and re-configurable scenarios. several numerical simulations are proposed which corroborate the validity and effectiveness of the propounded approach. As the first example, a directional heat bending device was designed that despite of its deflection angle, only a constant TNM can be used. In addition, we have used the TNM as a mean to obtain arbitrary shape heat concentrators with homogenous materials. It is shown that by utilizing a TNM, one can freely change the geometry of the concentrator and still use the same material. Finally, a thermal omnidirectional cloak is also proposed that is capable of guiding the thermal distributions in a manner that they become undetectable from an outside observer.  
\bibliography{achemso-demo}

\providecommand{\latin}[1]{#1}
\makeatletter
\providecommand{\doi}
  {\begingroup\let\do\@makeother\dospecials
  \catcode`\{=1 \catcode`\}=2 \doi@aux}
\providecommand{\doi@aux}[1]{\endgroup\texttt{#1}}
\makeatother
\providecommand*\mcitethebibliography{\thebibliography}
\csname @ifundefined\endcsname{endmcitethebibliography}
  {\let\endmcitethebibliography\endthebibliography}{}
\begin{mcitethebibliography}{28}
\providecommand*\natexlab[1]{#1}
\providecommand*\mciteSetBstSublistMode[1]{}
\providecommand*\mciteSetBstMaxWidthForm[2]{}
\providecommand*\mciteBstWouldAddEndPuncttrue
  {\def\EndOfBibitem{\unskip.}}
\providecommand*\mciteBstWouldAddEndPunctfalse
  {\let\EndOfBibitem\relax}
\providecommand*\mciteSetBstMidEndSepPunct[3]{}
\providecommand*\mciteSetBstSublistLabelBeginEnd[3]{}
\providecommand*\EndOfBibitem{}
\mciteSetBstSublistMode{f}
\mciteSetBstMaxWidthForm{subitem}{(\alph{mcitesubitemcount})}
\mciteSetBstSublistLabelBeginEnd
  {\mcitemaxwidthsubitemform\space}
  {\relax}
  {\relax}

\bibitem[Guenneau \latin{et~al.}(2012)Guenneau, Amra, and
  Veynante]{guenneau2012transformation}
Guenneau,~S.; Amra,~C.; Veynante,~D. Transformation thermodynamics: cloaking
  and concentrating heat flux. \emph{Optics Express} \textbf{2012}, \emph{20},
  8207--8218\relax
\mciteBstWouldAddEndPuncttrue
\mciteSetBstMidEndSepPunct{\mcitedefaultmidpunct}
{\mcitedefaultendpunct}{\mcitedefaultseppunct}\relax
\EndOfBibitem
\bibitem[Schittny \latin{et~al.}(2013)Schittny, Kadic, Guenneau, and
  Wegener]{schittny2013experiments}
Schittny,~R.; Kadic,~M.; Guenneau,~S.; Wegener,~M. Experiments on
  transformation thermodynamics: molding the flow of heat. \emph{Physical
  review letters} \textbf{2013}, \emph{110}, 195901\relax
\mciteBstWouldAddEndPuncttrue
\mciteSetBstMidEndSepPunct{\mcitedefaultmidpunct}
{\mcitedefaultendpunct}{\mcitedefaultseppunct}\relax
\EndOfBibitem
\bibitem[Pendry \latin{et~al.}(2006)Pendry, Schurig, and
  Smith]{pendry2006controlling}
Pendry,~J.~B.; Schurig,~D.; Smith,~D.~R. Controlling electromagnetic fields.
  \emph{science} \textbf{2006}, \emph{312}, 1780--1782\relax
\mciteBstWouldAddEndPuncttrue
\mciteSetBstMidEndSepPunct{\mcitedefaultmidpunct}
{\mcitedefaultendpunct}{\mcitedefaultseppunct}\relax
\EndOfBibitem
\bibitem[Liu \latin{et~al.}(2009)Liu, Ji, Mock, Chin, Cui, and
  Smith]{liu2009broadband}
Liu,~R.; Ji,~C.; Mock,~J.; Chin,~J.; Cui,~T.; Smith,~D. Broadband ground-plane
  cloak. \emph{Science} \textbf{2009}, \emph{323}, 366--369\relax
\mciteBstWouldAddEndPuncttrue
\mciteSetBstMidEndSepPunct{\mcitedefaultmidpunct}
{\mcitedefaultendpunct}{\mcitedefaultseppunct}\relax
\EndOfBibitem
\bibitem[Xu and Chen(2015)Xu, and Chen]{xu2015conformal}
Xu,~L.; Chen,~H. Conformal transformation optics. \emph{Nature Photonics}
  \textbf{2015}, \emph{9}, 15\relax
\mciteBstWouldAddEndPuncttrue
\mciteSetBstMidEndSepPunct{\mcitedefaultmidpunct}
{\mcitedefaultendpunct}{\mcitedefaultseppunct}\relax
\EndOfBibitem
\bibitem[Fakheri \latin{et~al.}(2017)Fakheri, Barati, and
  Abdolali]{fakheri2017carpet}
Fakheri,~M.~H.; Barati,~H.; Abdolali,~A. Carpet cloak design for rough
  surfaces. \emph{Chinese Physics Letters} \textbf{2017}, \emph{34},
  084101\relax
\mciteBstWouldAddEndPuncttrue
\mciteSetBstMidEndSepPunct{\mcitedefaultmidpunct}
{\mcitedefaultendpunct}{\mcitedefaultseppunct}\relax
\EndOfBibitem
\bibitem[Tichit \latin{et~al.}(2014)Tichit, Burokur, and
  de~Lustrac]{tichit2014spiral}
Tichit,~P.-H.; Burokur,~S.~N.; de~Lustrac,~A. Spiral-like multi-beam emission
  via transformation electromagnetics. \emph{Journal of Applied Physics}
  \textbf{2014}, \emph{115}, 024901\relax
\mciteBstWouldAddEndPuncttrue
\mciteSetBstMidEndSepPunct{\mcitedefaultmidpunct}
{\mcitedefaultendpunct}{\mcitedefaultseppunct}\relax
\EndOfBibitem
\bibitem[Zhang \latin{et~al.}(2016)Zhang, Ding, Wo, Meng, and
  Wu]{zhang2016experimental}
Zhang,~K.; Ding,~X.; Wo,~D.; Meng,~F.; Wu,~Q. Experimental validation of
  ultra-thin metalenses for N-beam emissions based on transformation optics.
  \emph{Applied Physics Letters} \textbf{2016}, \emph{108}, 053508\relax
\mciteBstWouldAddEndPuncttrue
\mciteSetBstMidEndSepPunct{\mcitedefaultmidpunct}
{\mcitedefaultendpunct}{\mcitedefaultseppunct}\relax
\EndOfBibitem
\bibitem[Jiang \latin{et~al.}(2012)Jiang, Gregory, and
  Werner]{jiang2012broadband}
Jiang,~Z.~H.; Gregory,~M.~D.; Werner,~D.~H. Broadband high directivity
  multibeam emission through transformation optics-enabled metamaterial lenses.
  \emph{IEEE Transactions on Antennas and Propagation} \textbf{2012},
  \emph{60}, 5063--5074\relax
\mciteBstWouldAddEndPuncttrue
\mciteSetBstMidEndSepPunct{\mcitedefaultmidpunct}
{\mcitedefaultendpunct}{\mcitedefaultseppunct}\relax
\EndOfBibitem
\bibitem[Ashrafian \latin{et~al.}(2019)Ashrafian, Fakheri, and
  Abdolali]{ashrafian2019space}
Ashrafian,~A.; Fakheri,~M.~H.; Abdolali,~A. Space wave channeling enabled by
  conformal transformation optics. \emph{JOSA B} \textbf{2019}, \emph{36},
  2499--2507\relax
\mciteBstWouldAddEndPuncttrue
\mciteSetBstMidEndSepPunct{\mcitedefaultmidpunct}
{\mcitedefaultendpunct}{\mcitedefaultseppunct}\relax
\EndOfBibitem
\bibitem[Barati \latin{et~al.}(2019)Barati, Fakheri, and
  Abdolali]{barati2019exploiting}
Barati,~H.; Fakheri,~M.~H.; Abdolali,~A. Exploiting transformation optics for
  arbitrary manipulation of antenna radiation pattern. \emph{IET Microwaves,
  Antennas \& Propagation} \textbf{2019}, \relax
\mciteBstWouldAddEndPunctfalse
\mciteSetBstMidEndSepPunct{\mcitedefaultmidpunct}
{}{\mcitedefaultseppunct}\relax
\EndOfBibitem
\bibitem[Rahm \latin{et~al.}(2008)Rahm, Schurig, Roberts, Cummer, Smith, and
  Pendry]{rahm2008design}
Rahm,~M.; Schurig,~D.; Roberts,~D.~A.; Cummer,~S.~A.; Smith,~D.~R.;
  Pendry,~J.~B. Design of electromagnetic cloaks and concentrators using
  form-invariant coordinate transformations of Maxwell’s equations.
  \emph{Photonics and Nanostructures-fundamentals and Applications}
  \textbf{2008}, \emph{6}, 87--95\relax
\mciteBstWouldAddEndPuncttrue
\mciteSetBstMidEndSepPunct{\mcitedefaultmidpunct}
{\mcitedefaultendpunct}{\mcitedefaultseppunct}\relax
\EndOfBibitem
\bibitem[Yang \latin{et~al.}(2009)Yang, Huang, Yang, Xiao, and
  Peng]{yang2009metamaterial}
Yang,~J.; Huang,~M.; Yang,~C.; Xiao,~Z.; Peng,~J. Metamaterial electromagnetic
  concentrators with arbitrary geometries. \emph{Optics Express} \textbf{2009},
  \emph{17}, 19656--19661\relax
\mciteBstWouldAddEndPuncttrue
\mciteSetBstMidEndSepPunct{\mcitedefaultmidpunct}
{\mcitedefaultendpunct}{\mcitedefaultseppunct}\relax
\EndOfBibitem
\bibitem[Sadeghi \latin{et~al.}(2015)Sadeghi, Li, Xu, Hou, and
  Chen]{sadeghi2015transformation}
Sadeghi,~M.; Li,~S.; Xu,~L.; Hou,~B.; Chen,~H. Transformation optics with
  Fabry-P{\'e}rot resonances. \emph{Scientific reports} \textbf{2015},
  \emph{5}, 8680\relax
\mciteBstWouldAddEndPuncttrue
\mciteSetBstMidEndSepPunct{\mcitedefaultmidpunct}
{\mcitedefaultendpunct}{\mcitedefaultseppunct}\relax
\EndOfBibitem
\bibitem[Zhao \latin{et~al.}(2018)Zhao, Xu, Cai, Liu, and
  Chen]{zhao2018feasible}
Zhao,~P.-F.; Xu,~L.; Cai,~G.-X.; Liu,~N.; Chen,~H.-Y. A feasible approach to
  field concentrators of arbitrary shapes. \emph{Frontiers of Physics}
  \textbf{2018}, \emph{13}, 134205\relax
\mciteBstWouldAddEndPuncttrue
\mciteSetBstMidEndSepPunct{\mcitedefaultmidpunct}
{\mcitedefaultendpunct}{\mcitedefaultseppunct}\relax
\EndOfBibitem
\bibitem[Yang \latin{et~al.}(2019)Yang, Huang, Yang, Li, Mao, and
  Li]{yang2019arbitrarily}
Yang,~C.~F.; Huang,~M.; Yang,~J.~J.; Li,~T.~H.; Mao,~F.~C.; Li,~P. Arbitrarily
  shaped homogeneous concentrator and its layered realization. \emph{Optics
  Communications} \textbf{2019}, \emph{435}, 150--158\relax
\mciteBstWouldAddEndPuncttrue
\mciteSetBstMidEndSepPunct{\mcitedefaultmidpunct}
{\mcitedefaultendpunct}{\mcitedefaultseppunct}\relax
\EndOfBibitem
\bibitem[Zhou \latin{et~al.}(2018)Zhou, Xu, Zhang, Wu, Li, and
  Chen]{zhou2018perfect}
Zhou,~M.-Y.; Xu,~L.; Zhang,~L.-C.; Wu,~J.; Li,~Y.-B.; Chen,~H.-Y. Perfect
  invisibility concentrator with simplified material parameters.
  \emph{Frontiers of Physics} \textbf{2018}, \emph{13}, 134101\relax
\mciteBstWouldAddEndPuncttrue
\mciteSetBstMidEndSepPunct{\mcitedefaultmidpunct}
{\mcitedefaultendpunct}{\mcitedefaultseppunct}\relax
\EndOfBibitem
\bibitem[Abdolali \latin{et~al.}(2019)Abdolali, Sedeh, and
  Fakheri]{abdolali2019geometry}
Abdolali,~A.; Sedeh,~H.~B.; Fakheri,~M.~H. Geometry free materials enabled by
  transformation optics for enhancing the intensity of electromagnetic waves in
  an arbitrary domain. \emph{arXiv preprint arXiv:1902.08254} \textbf{2019},
  \relax
\mciteBstWouldAddEndPunctfalse
\mciteSetBstMidEndSepPunct{\mcitedefaultmidpunct}
{}{\mcitedefaultseppunct}\relax
\EndOfBibitem
\bibitem[Barati \latin{et~al.}(2018)Barati, Fakheri, and
  Abdolali]{barati2018experimental}
Barati,~H.; Fakheri,~M.; Abdolali,~A. Experimental demonstration of
  metamaterial-assisted antenna beam deflection through folded transformation
  optics. \emph{Journal of Optics} \textbf{2018}, \emph{20}, 085101\relax
\mciteBstWouldAddEndPuncttrue
\mciteSetBstMidEndSepPunct{\mcitedefaultmidpunct}
{\mcitedefaultendpunct}{\mcitedefaultseppunct}\relax
\EndOfBibitem
\bibitem[Sedeh \latin{et~al.}(2019)Sedeh, Fakheri, and
  Abdolali]{sedeh2019advanced}
Sedeh,~H.~B.; Fakheri,~M.~H.; Abdolali,~A. Advanced synthesis of meta-antenna
  radiation pattern enabled by transformation optics. \emph{Journal of Optics}
  \textbf{2019}, \emph{21}, 045108\relax
\mciteBstWouldAddEndPuncttrue
\mciteSetBstMidEndSepPunct{\mcitedefaultmidpunct}
{\mcitedefaultendpunct}{\mcitedefaultseppunct}\relax
\EndOfBibitem
\bibitem[Kwon and Werner(2008)Kwon, and Werner]{kwon2008polarization}
Kwon,~D.-H.; Werner,~D.~H. Polarization splitter and polarization rotator
  designs based on transformation optics. \emph{Optics Express} \textbf{2008},
  \emph{16}, 18731--18738\relax
\mciteBstWouldAddEndPuncttrue
\mciteSetBstMidEndSepPunct{\mcitedefaultmidpunct}
{\mcitedefaultendpunct}{\mcitedefaultseppunct}\relax
\EndOfBibitem
\bibitem[Yu \latin{et~al.}(2011)Yu, Lin, Zhang, Yu, Yu, and Su]{yu2011design}
Yu,~G.-x.; Lin,~Y.-f.; Zhang,~G.-q.; Yu,~Z.; Yu,~L.-l.; Su,~J. Design of
  square-shaped heat flux cloaks and concentrators using method of coordinate
  transformation. \emph{Frontiers of Physics in China} \textbf{2011}, \emph{6},
  70--73\relax
\mciteBstWouldAddEndPuncttrue
\mciteSetBstMidEndSepPunct{\mcitedefaultmidpunct}
{\mcitedefaultendpunct}{\mcitedefaultseppunct}\relax
\EndOfBibitem
\bibitem[Hu \latin{et~al.}(2016)Hu, Zhou, Shu, Xie, Ma, and
  Luo]{hu2016directional}
Hu,~R.; Zhou,~S.; Shu,~W.; Xie,~B.; Ma,~Y.; Luo,~X. Directional heat transport
  through thermal reflection meta-device. \emph{AIP Advances} \textbf{2016},
  \emph{6}, 125111\relax
\mciteBstWouldAddEndPuncttrue
\mciteSetBstMidEndSepPunct{\mcitedefaultmidpunct}
{\mcitedefaultendpunct}{\mcitedefaultseppunct}\relax
\EndOfBibitem
\bibitem[Farhat \latin{et~al.}(2016)Farhat, Chen, Guenneau, and
  Enoch]{farhat2016transformation}
Farhat,~M.; Chen,~P.-Y.; Guenneau,~S.; Enoch,~S. \emph{Transformation wave
  physics: electromagnetics, elastodynamics, and thermodynamics}; CRC Press,
  2016\relax
\mciteBstWouldAddEndPuncttrue
\mciteSetBstMidEndSepPunct{\mcitedefaultmidpunct}
{\mcitedefaultendpunct}{\mcitedefaultseppunct}\relax
\EndOfBibitem
\bibitem[Sun \latin{et~al.}(2019)Sun, Liu, Yang, Chen, and He]{sun2019thermal}
Sun,~F.; Liu,~Y.; Yang,~Y.; Chen,~Z.; He,~S. Thermal surface transformation and
  its applications to heat flux manipulations. \emph{Optics Express}
  \textbf{2019}, \emph{27}, 33757--33767\relax
\mciteBstWouldAddEndPuncttrue
\mciteSetBstMidEndSepPunct{\mcitedefaultmidpunct}
{\mcitedefaultendpunct}{\mcitedefaultseppunct}\relax
\EndOfBibitem
\bibitem[Liu \latin{et~al.}(2018)Liu, Sun, and He]{liu2018fast}
Liu,~Y.; Sun,~F.; He,~S. Fast Adaptive Thermal Buffering by a Passive Open
  Shell Based on Transformation Thermodynamics. \emph{Advanced Theory and
  Simulations} \textbf{2018}, \emph{1}, 1800026\relax
\mciteBstWouldAddEndPuncttrue
\mciteSetBstMidEndSepPunct{\mcitedefaultmidpunct}
{\mcitedefaultendpunct}{\mcitedefaultseppunct}\relax
\EndOfBibitem
\bibitem[Yang \latin{et~al.}(2014)Yang, Vemuri, and
  Bandaru]{yang2014experimental}
Yang,~T.; Vemuri,~K.~P.; Bandaru,~P.~R. Experimental evidence for the bending
  of heat flux in a thermal metamaterial. \emph{Applied Physics Letters}
  \textbf{2014}, \emph{105}, 083908\relax
\mciteBstWouldAddEndPuncttrue
\mciteSetBstMidEndSepPunct{\mcitedefaultmidpunct}
{\mcitedefaultendpunct}{\mcitedefaultseppunct}\relax
\EndOfBibitem
\end{mcitethebibliography}

\end{document}